# The Evolution of Multicomponent Systems at High Pressures: VI.
# The Thermodynamic Stability of the Hydrogen-Carbon System:
# The Genesis of Hydrocarbons and the Origin of Petroleum.

**J. F. Kenney,** (JFK@alum.MIT.edu)*
Gas Resources Corporation, 11811 North Parkway, fl. 5, Houston, TX 77060, Houston, U.S.A.
Russian Academy of Sciences - Joint Institute of Earth Physics, Bolshaya Gruzinskaya 10, 123.810 Moscow, Russia;

**Vladimir A. Kutcherov**
Russian State University of Oil and Gas, Leninski prospect 65, 117917 Moscow, Russia.

**Nikolai A. Bendeliani**, **Vladimir A. Alekseev**
Russian Academy of Sciences - Institute for High Pressure Physics, 142092 Troitsk, Moscow Region, Russia.

**Abstract:**

The spontaneous genesis of hydrocarbons which comprise natural petroleum have been analyzed by chemical thermodynamic stability theory. The constraints imposed upon chemical evolution by the second law of thermodynamics are briefly reviewed; and the effective prohibition of transformation, in the regime of temperatures and pressures characteristic of the near-surface crust of the Earth, of biological molecules into hydrocarbon molecules heavier than methane is recognized.

For the theoretical analysis of this phenomenon, a general, first-principles equation of state has been developed by extending scaled particle theory (SPT) and by using the technique of the factored partition function of the Simplified Perturbed Hard

---

* author for communications.

Chain Theory (SPHCT). The chemical potentials, and the respective thermodynamic Affinity, have been calculated for typical components of the hydrogen-carbon (H-C) system over a range pressures between 1-100 kbar, and at temperatures consistent with those of the depths of the Earth at such pressures. The theoretical analyses establish that the normal alkanes, the homologous hydrocarbon group of lowest chemical potential, evolve only at pressures greater than approximately thirty kbar, excepting only the lightest, methane. The pressure of thirty kbar corresponds to depths of approximately 100 km.

For experimental verification of the predictions of the theoretical analysis, special high-pressure apparatus has been designed which permits investigations at pressures to 50 kbar and temperatures to 1500°C, and which also allows rapid cooling while maintaining high pressures. The high-pressure genesis of petroleum hydrocarbons has been demonstrated using *only* the solid reagents iron oxide, FeO, and marble, $CaCO_3$, 99.9% pure and wet with triple-distilled water.

Natural petroleum is a hydrogen-carbon [H-C] system, in distinctly non-equilibrium states, composed of mixtures of highly reduced, hydrocarbon molecules, all of very high chemical potential, most in the liquid phase. As such, the phenomenon of the terrestrial existence of natural petroleum in the near-surface crust of the Earth has presented several challenges, most of which have remained unresolved until recently. The primary scientific problem of petroleum has been the existence and genesis of the individual hydrocarbon molecules themselves: how, and under what thermodynamic conditions, can such highly-reduced molecules of high chemical potential evolve.

The scientific problem of the genesis of hydrocarbons of natural petroleum, and consequentially of the origin of natural petroleum deposits, has regrettably been one too much neglected by competent physicists and chemists; the subject has been obscured by diverse, unscientific hypotheses, typically connected with the rococo hypothesis(1) that highly-reduced hydrocarbon molecules of high chemical potentials might somehow evolve from highly-oxidized biotic molecules of low chemical potential. The scientific problem of the spontaneous evolution of the hydrocarbon molecules comprising natural petroleum is one of chemical thermodynamic stability theory. This problem does *not* involve the properties of rocks where petroleum might be found, nor of microorganisms observed in crude oil.

This paper is organized into five parts. The first section reviews briefly the formalism of modern thermodynamic stability theory, the theoretical framework for the analysis of the genesis of hydrocarbons and the H-C system, - as similarly for any system.

The second section examines, applying the constraints of thermodynamics, the notion that hydrocarbons might evolve spontaneously from biological molecules. Here are described the spectra of chemical potentials of hydrocarbon molecules, particularly the naturally-occurring ones present in petroleum. Interpretation of the significance of the relative differences between the chemical potentials of the hydrocarbon system and those of biological molecules, applying the dictates of thermodynamic stability theory, disposes of any hypothesis of an origin for hydrocarbon molecules from biological matter, excepting only the lightest, methane.

In the third section is described a first-principles, statistical mechanical formalism, developed from an extended representation of scaled particle theory appropriate for mixtures of aspherical, hard-body molecules, combined with a mean-field representation of the long-range, attractive component of the intermolecular potential.

In the fourth section, the thermodynamic Affinity developed using this formalism establishes that the hydrocarbon molecules peculiar to natural petroleum are high-pressure polymorphs of the H-C system, similarly as diamond and lonsdalite are to graphite for the elemental carbon system, and evolve only in thermodynamic regimes of pressures greater than 25-50 kbar.

The fifth section reports the experimental results obtained using equipment specially-designed to test the predictions of the previous sections. Application of pressures to 50 kbar and temperatures to 1500°C upon solid (and obviously abiotic) $CaCO_3$ and FeO, wet with triple-distilled water, all in the absence of any initial hydrocarbon or biotic molecules, evolves the suite of petroleum fluids: methane, ethane, propane, butane, pentane, hexane, branched isomers of those compounds, and the lightest of the n-alkene series.

**1.   Thermodynamic stability and the evolution of multicomponent systems.**

Central to any analysis of chemical stability is the thermodynamic Affinity, $A(\{m_i\})$, which determines the direction of evolution of a system in accordance with the second law of thermodynamics, as expressed by De Donder's inequality, $dQ' = A d\chi \geq 0$.(2) The Affinity of an $n$-component, multiphase system of $p$ phases involving $r$ chemical reactions is given as:

$$A = \sum_{r=1}^{r} A_r = -\sum_{r=1}^{r}\sum_{a=1}^{p}\sum_{i=1}^{n} n_{i,r} m_i^a\left(p, T, \{n_j\}\right). \qquad (1)$$

in which $m_{i,r}$ and $n_{i,r}$ are, respectively, the chemical potential and stoichiometric coefficients of the $i$-th component in the $r$-th reaction; $a$ designates the respective phase.

The second law states that the internal production of entropy is always positive for every spontaneous transformation. Therefore, the thermodynamic Affinity, (1), must always be positive, and the direction of evolution of any system must always obey the inequalities:

$$dS_{int} = \begin{cases} \dfrac{1}{T}\sum_r A_r d\mathbf{x}_r = -\dfrac{1}{T}\sum_{r=1}^{r}\sum_{a=1}^{p}\sum_{i=1}^{n} n_{i,r} \mathbf{m}_i^a\left(p,T,\{n_j\}\right) d\mathbf{x}_r \\ \sum_k F_k dX_k \end{cases} \geq 0. \quad (2)$$

The inequalities (2) express the irreversibility of spontaneous transitions, and state, that: for a spontaneous evolution of a system from any state, A, to any other state, B, the free enthalpy of state B must be less than that of state A; and that at no point between the two may the free enthalpy be greater than that of state A, or less than that of state B.

The sum of the products on the second line of inequality in (2), of the thermodynamic Affinities and the differential of the variables of extent, $d\mathbf{x}_r$, is always positive; and the circumstance for which the change of internal entropy is zero defines equilibrium, from which there is no spontaneous evolution. This is De Donder's inequality.

The sum of products on the second line of inequality in (2) deserves particular note. In the second line of (2), $F_k$ and $dX_k$ are, respectively, general thermodynamic forces and flows, and subsume Newton's rule, $\vec{F}=m\vec{a}$, as a special case.(3, 4) The expression in the second line of (2) states further that, for any circumstance for which the Affinity does not vanish, there exists a generalized thermodynamic force which drives the system toward equilibrium. The constraints of this expression assure that an apple, having disconnected from its bough, does not fall, say, half way to the ground and there stop (a phenomenon not prohibited by the first law), but must continue to fall until the ground. These constraints force a chemically reactive system to evolve always toward the state of lowest thermodynamic Affinity

Thus, the evolution of a chemically reactive, multicomponent system may be determined at any temperature, pressure, or composition whenever the chemical potentials of its components are known. To ascertain the thermodynamic regime of the spontaneous evolution of hydrocarbons their chemical potentials must be determined.

No consideration has been given, in the foregoing discussion of chemical thermodynamic stability, to the rate of increase of the variables of extent, $d\mathbf{x}_r$. Such is the subject of chemical kinetics, not stability theory. The rate at which a reaction might occur cannot alter its direction, as determined by the second law of thermodynamics, - otherwise the second law would not exist. The evolution of a system can admit intermediate states, in which one (or more) intermediate product might possess a chemical potential considerably greater than that of any of the initial reagents. The presence of a selected catalyst can enhance a fast, reaction; and, if the system is removed rapidly from thermodynamic environment at which such reactions proceeding to the final state occur, the intermediate product(s) can be separated. The petrochemical industry routinely operates such processes. However, such complex, industrial processes are not mimicked spontaneously in the natural world.

2. **The thermodynamic energy spectrum of the hydrogen-carbon [H-C] system; and the effective prohibition of low-pressure genesis of hydrocarbons.**

The thermodynamic energy spectrum of the chemical potentials [molar Gibbs energies of formation, $\Delta G_f$] of the H-C system at standard temperature and pressure (STP: 298.15 K; 1 atm) are available from tables of chemical data.(5) The chemical potentials of naturally occurring members of the of the H-C system at STP are shown graphically in Fig. 1. Examination of the energy spectrum of these chemical potentials of the H-C system establishes at once that, at STP, the chemical potentials of the entire hydrocarbon system are remarkable for both their characteristic *increase* with degree of polymerization as well as their linear, and almost constant, magnitude of such increase with carbon number. With increasing polymerization, the n-alkane molecules manifest increased chemical potential of very approximately 2.2 kcal per added carbon atom, or $CH_2$ unit. (There exist also branched isomers whose chemical potentials differ from such of the normal configuration by, typically, 2-4%.) Such increase in chemical potential with increased degree of polymerization contrasts strongly with the thermodynamic spectrum of the highly oxidized, biotic carbon ("organic") compounds of the hydrogen-carbon-oxygen [H-C-O] system, which manifest consistently *decreasing* chemical potentials with increasing polymerization. This latter property allows the high degree of polymerization and complexity of the biotic compounds.

Examination of the H-C-O system of oxidized carbon compounds establishes that the chemical potentials of almost all biotic compounds lie far below that of methane, the least energetic of the reduced hydrocarbon compounds, typically by several hundred kcal/mol. Although there exist biotic molecules of unusually high chemical potential, such as β-carotene ($C_{40}H_{56}$), vitamin-D ($C_{38}H_{44}O$), and some of the pheromone hormones, such compounds are relatively rare by abundance. They are produced by biological systems only when the producing entity is alive, and at formidable metabolic cost to the producing entity; and the production ceases with the death of the entity. Such compounds are not decomposition products of other biotic compounds, and are labile and themselves decompose rapidly. For these foregoing reasons, such compounds cannot be considered relevant to the subject of the origin of natural petroleum.

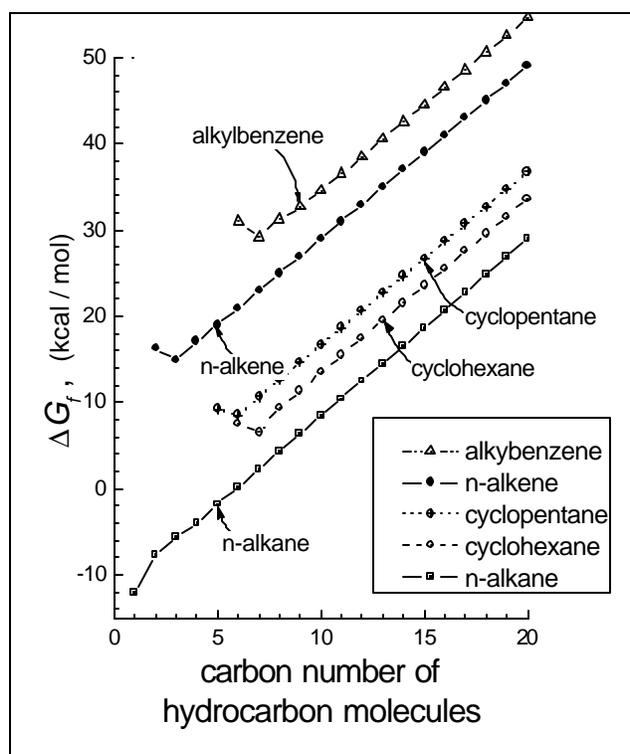

**Fig. 1 Molar Gibbs energies of formation, $\Delta G_f$, of the naturally occurring hydrocarbons, at STP.(5)**

The properties of the thermodynamic energy spectrum of the H-C and H-C-O systems, together with the constraints of the second law, (2), establish three crucial properties of natural petroleum:

(**1.**) **The H-C system which constitutes natural petroleum is a metastable one in a very non-equilibrium state. At low pressures, all heavier hydrocarbon molecules are thermodynamically unstable against decomposition into methane and carbon, - as is similarly diamond into graphite.**

(**2.**) **Methane does *not* polymerize into heavy hydrocarbon molecules at low pressures, at *any* temperature. Contrarily, increasing temperature (at low pressures) must increase the rate of decomposition of heavier hydrocarbons into methane and carbon.**

(**3.**) **Any hydrocarbon compound generated at low pressures, heavier than methane, would be unstable and driven to the stable equilibrium state of methane and carbon.**

These conclusions have been amply demonstrated by a century of refinery engineering practice. The third conclusion has been demonstrated by many unsuccessful laboratory attempts to convert biotic molecules into hydrocarbons heavier than methane.

There are three generic chemical processes which deserve specific consideration: the "charcoal burner's reaction," the "bean-eater's reaction," and the "octane-enhanced bean-eater's reaction." All describe limited reactions by which a highly oxidized biotic molecule can react to produce elemental carbon when "carried" by a more thorough oxidation process. In both the following examples, the simple carbohydrate, sugar $C_6H_{12}O_6$, is used as a typical biotic reagent; the same reasoning and results hold also for any of the highly-oxidized biotic compounds.

The "charcoal burner's reaction" is: $C_6H_{12}O_6 \rightarrow 6C + 6H_2O$. The chemical potentials at STP for the simple carbohydrate $C_6H_{12}O_6$, and water vapor at STP are, respectively in kcal/mol: 218.720; -54.636. The thermodynamic Affinity for the "charcoal burner's reaction" to produce amorphous carbon, or graphite, is, accordingly, 109.10 kcal. Therefore, the genesis of coal from biological detritus in an oxygen-poor environment is permitted by the second law. However, the thermodynamic Affinity for the "charcoal burner's reaction" to produce diamond is 105.02 kcal, which quantity is also positive, and therefore *not* immediately prohibited by the second law as expressed solely by de Donder's inequality, the first of equations (2). Nonetheless, no charcoal burner ever scrabbles through his ashes hoping to find diamonds. Such reasonable behavior demonstrates an effective appreciation of the dictates of the second law as expressed by the second of equations (2). In this case, the generalized force is the difference in thermodynamic Affinity between the reactions for graphite and diamond, respectively, $dF = dA/T$, which, in the regime of temperatures and pressures of the near-surface crust of the Earth, assures always spontaneous genesis of graphite, but never of diamond. Similarly, for reactions involving hydrocarbons heavier than methane, the same generalized force, $dA/T$, always drives the system toward the state of lowest free enthalpy, i.e., methane plus free carbon.

The "bean-eater's reaction" is: $C_6H_{12}O_6 \rightarrow 3CH_4 + 3CO_2$. The chemical potentials at STP for methane, and carbon dioxide are, respectively in kcal/mol: –12.130; and -94.260. The thermodynamic Affinity for this reaction is accordingly 100.42 kcal/mol, and therefore permitted by the second law. Indeed, reactions of the type are typical of those by which methane is produced in swamps, sewers, and the bowels of herbivores.

The "octane-enhanced bean-eater's reaction" is: $C_6H_{12}O_6 \rightarrow 1/8 C_8H_{18} + 7/8 H_2 + 2CH_4 + 3CO_2$. Since the chemical potential at STP of n-octane is 4.290 kcal/mol, and that of molecular hydrogen is zero, the thermodynamic Affinity for the "octane-enhanced bean-eater's reaction" is $A = (100.42 - 12.130 - 4.290/8) = 87.70$ kcal/mol, still positive and thereby not prohibited outright by the constraints of De Donder's inequality. However, no biochemical investigation has ever observed a molecule of any hydrocarbon heavier than methane resulting from the decomposition of biological detritus. After a meal of, e.g., Boston baked beans, one does experience biogenic methane, - but *not* "biogenic" octane. No such process produces heavier hydrocarbons, for such process would involve effectively a reaction of low-pressure methane polymerization, similarly as the effective prohibition of the evolution of diamonds by the "charcoal burner's reaction." In the previous section was described the industrial technique by which useful intermediate products can be obtained by controlling the reaction process. The Fischer-Tropsch process uses reactions essentially identical to the "octane-enhanced bean-eater's reaction" to generate liquid petroleum fuels from the combustion of coal, wood, or other biotic matter. However, the highly-controlled, industrial Fischer-Tropsch process does not produce, uncontrolled and spontaneously, the commonly observed, large accumulations of natural petroleum.

The foregoing properties of natural petroleum, and the effective prohibition by the second law of thermodynamics of its spontaneous genesis from highly-oxidized biological molecules of low chemical potentials, were clearly understood in the second half of the 19th century by chemists and thermodynamicists, such as Berthelot, and later confirmed by others, including Sokolov, Biasson, and Mendeleev. However, the problem of how, and in what regime of temperature and pressure, hydrogen and carbon combine to form the particular H-C system manifested by natural petroleum, remained. The resolution of this problem had to wait a century for the development of modern atomic and molecular theory, quantum statistical mechanics, and many-body theory. This problem has now been resolved theoretically by determination of the chemical potentials and the thermodynamic Affinity of the H-C system, using modern quantum statistical mechanics, and has also now been demonstrated experimentally with specially designed high-pressure apparatus.

### 3. Calculation of the thermodynamic Affinity using scaled particle theory [SPT] and the formalism of the Simplified Perturbed Hard Chain Theory [SPHCT].

In order to calculate the thermodynamic Affinity of a distribution of compounds of the hydrogen-carbon (H-C) system in general regimes of temperature and pressure, one must use a rigorous mathematical formalism developed from first-principles, statistical mechanical argument. No approximate, or interpolated, formalism developed for the low-pressure regime can suffice. A sufficiently rigorous formalism has been developed by extending scaled particle theory [SPT] equation of state of Pavlícek, Nezbeda, and Boublík,(6, 7) for mixtures of convex, hard-body systems, combined with the formalism of the Simplified Perturbed Hard Chain Theory (SPHCT).(8)

Following the procedure enunciated originally by Bogolyubov,(9) and developed further by Feynmann(10) and Yukhnovski,(11) a factored partition function is used which employs a reference system: $Q = Q^{\text{ref}} Q^{\text{vdW}}$. The reference system employed is that of the hard-body fluid as described rigorously by scaled particle theory [SPT].(12-15) The description of the hard-body fluid by scaled particle theory is one of the few *exactly-solvable* problems in statistical mechanics. This property is especially valuable because the thermodynamic evolution of a system at high pressures is determined almost entirely by the variables of its components that are obtained from the reference system.

For mixtures of hard-body particles of different sizes and shapes, scaled particle theory generates the following analytical expression for the contribution to the pressure of the hard-core reference system:

$$p^{\text{ref}} = k_B T r \left[ 1 + \left( \frac{h}{(1-h)} + \frac{rs}{r(1-h)^2} + \frac{qs^2(1-2h)+5rsh^2}{3r(1-h)^3} \right) \right] = \left( p^{\text{IG}} + p^{\text{hc}} \right) \qquad (3)$$

in which the geometric compositional variables, *r*, *s*, and **h**, are defined by the Steiner-Kihara equations:

$$r = \boldsymbol{r}\sum_i x_i \tilde{R}_i, \quad q = \boldsymbol{r}\sum_i x_i \tilde{R}_i^2, \quad s = \boldsymbol{r}\sum_i x_i \tilde{S}_i, \quad \boldsymbol{u} = \sum_i x_i \tilde{V}_i, \quad \boldsymbol{h} = \boldsymbol{r}\sum_i x_i \tilde{V}_i = \boldsymbol{r}\boldsymbol{u} \quad \bigg\}.(4)$$

(A thorough discussion of the Steiner-Kihara parameters may be found in (16).) The following geometric functions are introduced: $\boldsymbol{a} = rs/3\boldsymbol{u}$, and $\boldsymbol{g} = <r^2>/<r>^2$. The geometric parameter $\boldsymbol{a}$ is the multi-component analogue of the Boublík parameter of asphericity for a single-component fluid, $\boldsymbol{a}^B = \tilde{R}\tilde{S}/(3\tilde{V})$, and may be interpreted as the system's weighted degree of asphericity; the parameter $\boldsymbol{g}$ has no analogue in a single-component fluid, for which it is always equal to unity; $\boldsymbol{g}$ might be interpreted as a parameter of interference which measures the degree of difference in the mean component dimensions of radii. When these definitions are used, the Boublík equation, (3), can be written in a simple form as,

$$p^{ref} = k_B T \boldsymbol{r}\left[1 + \frac{c_1\boldsymbol{h} + c_2\boldsymbol{h}^2 + c_3\boldsymbol{h}^3}{(1-\boldsymbol{h})^3}\right], \tag{5}$$

in which $c_1$, $c_2$, and $c_3$ are variables of composition which depend upon the combined geometries of the molecular components and their fractional abundances: $c_1 = 3\boldsymbol{a}+1$, $c_2 = 3\boldsymbol{a}(\boldsymbol{ag}-1)-2$, $c_3 = 1-\boldsymbol{a}(6\boldsymbol{ag}-5)$.

Similarly, the contribution of the reference system to the Free enthalpy may be written, as:

$$G^{ref} = Nk_B T\left[\sum_i x_i \ln\left(\frac{(n_i\boldsymbol{l}_i^3)}{V}\right) + \boldsymbol{h}\left(\frac{I + J\boldsymbol{h} + K\boldsymbol{h}^2}{(1-\boldsymbol{h})^3}\right) - c_3 \ln(1-\boldsymbol{h})\right], \tag{6}$$

in which $I = 2c_1-c_3$, $J = -1/2(3c_1-3c_2-5c_3)$, $K = 1/6(3c_1-3c_2-3c_3)$. When these identities are used, the contribution of the reference system to the pressure and the Free enthalpy become simplified functions of the packing-fraction, $\boldsymbol{h}$, and the geometric compositional variables, $\boldsymbol{a}$ and $\boldsymbol{g}$.

The contributions to the pressure and the chemical potentials from the long-range van der Waals component of the intermolecular potential are described using the formalism of the Simplified Perturbed Hard Chain Theory [SPHCT].(17-19) The SPHCT uses the mean-field technique,(20) of the Bethe-Peierls-Prigogine "lattice-gas" model, in which has been applied the shape-independent scattering formalism.(21) As demonstrated previously,(22) at elevated pressures, the pressure and chemical potential are dominated by their respective hard-core components, and the attractive component is several orders of magnitude smaller, and of little consequence. The representation of the attractive components of the pressure and the chemical potential used has been that developed for the SPHCT by Sandler,(23), Donohue, (19), Lee and Chao,(24) using the mixing rules of Kim.(25) The Prigogine shape $c$-factors used by the SPHCT are related to the Boublík geometric parameters such that:

$c_i = \left(1 + 3\tilde{R}_i \tilde{S}_i / \tilde{V}_i\right)$; and $V_i^* = \tilde{V}_i(1 + a_i)$. The values of $V_i^*$ and $a_i$ were taken from van Pelt *et al.*(26) The chemical potential of the *i*-th specie of a multi-component system is given by $\mu_i = \mu_i^\ominus + \mu_i^{ref} + \mu_i^{vdW}$, in which $\mu_i^\ominus$ represents the reference value of the chemical potential at standard temperature and pressure.

## 5. The evolution of the normal alkanes, ethane, hexane, and decane, from methane at high pressures.

At standard temperature and pressure, methane possesses the lowest chemical potential and is the only thermodynamically stable hydrocarbon. At low pressures, and all temperatures, all hydrocarbons are thermodynamically unstable relative to methane, or methane plus carbon (either graphite or amorphous carbon). At normal temperatures and pressures, the evolution of methane will dominate and effectively exhaust the H-C system of its elemental components. Because methane is the sole hydrocarbon specie which is thermodynamically stable at low pressures, the chemical Affinities of each of the heavier species have been calculated in comparison with methane. Accordingly, the chemical Affinity calculated for the thermodynamic stability of, for example, the methane ↔ (n-octane + hydrogen) system, is that for the reaction: $CH_4 \rightarrow 1/8\ C_8H_{18} + 7/8\ H_2$.

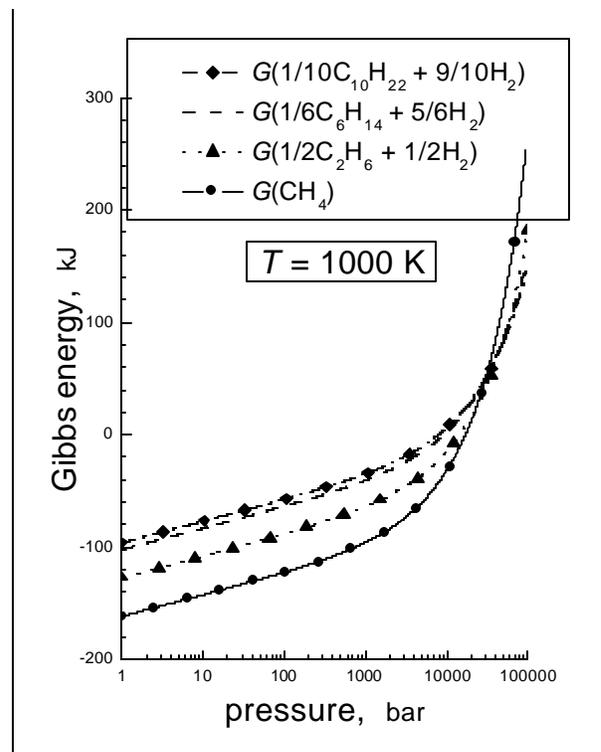

The chemical potentials of the hydrocarbon and methane molecules, and the resulting thermodynamic Affinities of the (methane ⇌ hydrocarbon + hydrogen) system, have been evaluated for the n-alkanes from methane through $C_{20}H_{42}$. In

**Fig. 2 Gibbs energies of methane and of the H-C system $[(1/n)C_nH_{2n+2} + (n-1)/nH_2]$.**

Fig. 2 are shown the Gibbs energies for the set of hydrocarbons methane ($CH_4$), and the n-alkanes, ethane (n-$C_2H_6$), hexane (n-$C_6H_{14}$), and decane (n-$C_{10}H_{22}$). These thermodynamic variables have been determined at pressures ranging from 1-100,000 bar and at the supercritical temperature 1000 K, which temperature corresponds conservatively to the geological regime characterized by the respective pressures of transition.

The values of the SPHCT parameters, *c*, *h*, and *Y*, for the individual compounds which have been used were taken from van Pelt *et al.*,(27) and the reference values of the chemical potentials of the pure component were taken from standard reference tables.(3)

The results of the analysis are shown graphically for the temperature 1000 K in Fig. 2. These results demonstrate clearly that all hydrocarbon molecules are chemically and thermodynamically unstable relative to methane at pressures less than approximately 25 kbar for the lightest, ethane, and 40 kbar for the heaviest n-alkane shown, decane.

The results of this analysis, shown graphically in Fig. 2, establish clearly the following:

**1.) With the exception of methane, heavier hydrocarbon molecules of higher chemical potentials are not generated spontaneously in the low-pressure regime of methane synthesis.**

**2.) All hydrocarbon molecules other than methane are high-pressure polymorphs of the H-C system and evolve spontaneously only at high pressures, greater than, at least, 25 kbar even under the most favorable circumstances.**

**3.) Contrary to experience of refinery operations conducted at low pressures, heavier alkanes are *not* unstable and do *not* necessarily decompose at elevated temperatures. Contrarily, at high pressures, methane transforms into the heavier alkanes, and the transformation processes are enhanced by elevated temperatures.**

The theoretical analyses reported here describe the high-pressure evolution of hydrocarbons under the *most favorable* chemical conditions. Therefore, although this analysis describes the thermodynamic stability of the H-C system, it does not explicitly do the same for the genesis of natural petroleum in the conditions of the depths of the Earth. The chemical conditions of the Earth, particularly near its surface, are oxidizing, *not* reducing; of the gases in the Earth's atmosphere and crust, hydrogen is significantly absent and methane a very minor constituent.

Although both methane and heavier hydrocarbons were present in the carbonaceous meteorites which participated in the accretion process of the formation of the Earth, such molecules were unlikely to have survived in their initial composition. The heat and impact which accompanied accretion would most likely have caused decomposition of heavier hydrocarbons and the release of methane. For both the theoretical analyses described in this section and the experimental investigations described in **section 6**, the conservative perspective has been taken that hydrocarbons evolve from the solid, abiotic carbon compounds and vestigial water present in the upper mantle of the Earth.

## 6. Experimental demonstration of hydrocarbon genesis, under thermodynamic conditions typical of the depths of the Earth.

Because the H-C system typical of petroleum is generated at high pressures, and exists only as a metastable mélange at laboratory pressures, special high-pressure apparatus has been designed which permits investigations at pressures to 50 kbar and temperatures to 1500°C, and which also allows rapid cooling while maintaining high pressures.(28) The importance of this latter ability cannot be overstated; for, in order to examine the spontaneous reaction products, the system must be rapidly quenched to "freeze in" their high-pressure, high-temperature distribution. Such mechanism is analogous to that which occurs during eruptive transport processes responsible for kimberlite ejecta, and for the stability and occurrence of diamonds in the crust of the Earth.

Experiments to demonstrate the high-pressure genesis of petroleum hydrocarbons have been carried out using *only* 99.9% pure, solid iron oxide, FeO, and marble, $CaCO_3$, wet with triple-distilled water. There were *no* biotic compounds or hydrocarbons admitted to the reaction chamber. The use of marble, instead of elemental carbon, was intentionally conservative. The initial carbon compound, $CaCO_3$, is more oxidized and of lower chemical potential. All of which rendered the system more resistant to the reduction of carbon to form heavy alkanes, than it would be under conditions of the mantle of the Earth. Although there has been observed igneous $CaCO_3$ (carbonatite) of mantle origin, carbon should be more reasonably expected to exist in the mantle of the Earth as an element in its dense phases: cubic (diamond), hexagonal (lonsdaleite), or random-close-pack (chaoite).

Pressure in the reaction cell, as described in (25) of volume 0.6 cm$^3$, was measured by a pressure gauge calibrated using data of the phase transitions of Bi, Tl, and PbTe. The cell was heated by a cylindrical graphite heater; its temperature was measured using a chromel-alumel thermocouple and was regulated within the range ±5°C. Both stainless steel and platinum reaction cells were used; all were constructed to prevent contamination by air and to provide impermeability during and after each experimental run.

The reaction cell was brought from 1 bar to 50 kbar gradually, at a rate of 2 kbar/min, and from room temperature to the elevated temperatures of investigation at the rate of 100 K/min. The cell and reaction chamber were held for at least an hour at each temperature for which measurements were taken in order to allow the H-C system to come to thermodynamic equilibrium. The samples were thenafter quenched rapidly at the rate of 700°C/sec to 50°C, and from 50°C to room temperature over several minutes, while maintaining the high pressure of investigation. The pressure was then reduced gradually to 1 bar at the rate of 1 kbar/min. The reaction cell was then gently heated to desorb the hydrocarbons for mass spectrometer analysis, using an HI-1201B mass spectrometer equipped with an automatic system of computerized spectrum registration. A specially-designed high-temperature gas probe allowed sampling the cell while maintaining its internal pressure.

At pressures below 10 kbar, no hydrocarbons heavier than methane were present. Hydrocarbon molecules began to evolve above 30 kbar. At 50 kbar and at the temperature of 1500°C, the system spontaneously evolved methane, ethane, n-propane, 2-methylpropane, 2,2-dimethylpropane, n-butane, 2-methylbutane, n-pentane, 2-methylpentane, n-hexane, and n-alkanes through $C_{10}H_{22}$, ethene, n-propene, n-butene, and n-pentene, in distributions characteristic of natural petroleum. The cumulative abundances of the subset of evolved hydrocarbons consisting of methane and n-alkanes through n-$C_6H_{14}$ are shown in Fig. 3 as functions of temperature. Methane (on the right scale) is present and of abundance approximately an order of magnitude greater than any single component of the heavier n-alkanes, although as a minor component of the total H-C system. That the extent of hydrocarbon evolution becomes relatively stable as a function of temperature above approximately 900°C, both for the absolute abundance of the individual hydrocarbon species as well as for their relative abundances, argues that the distributions observed represent thermodynamic equilibrium for the H-C system. That the evolved hydrocarbons remain stable over a range of temperatures increasing by more than 300 K demonstrates the third prediction of the theoretical analysis: **Hydrocarbon molecules heavier than methane do *not* decompose with increasing temperature in the high-pressure regime of their genesis**.

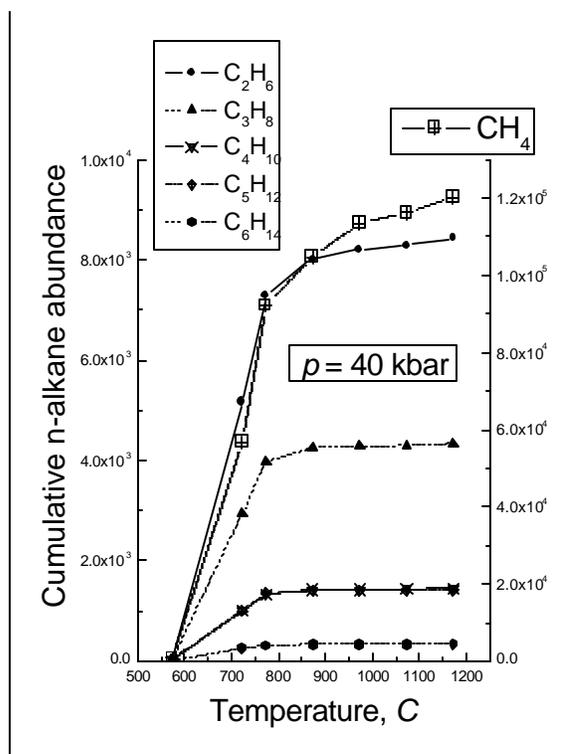

**Fig. 3 Cumulative abundances of n-alkanes through n-$C_6H_{14}$ on left ordinate, methane abundance on right, as functions of temperature at the pressure of 40 kbar. (Scales are in ppm.)**

### 7.   Discussion and conclusions.

The pressure of 30 kbar, at which the theoretical analyses of the **section 5** predicts that the H-C system must evolve ethane and heavier hydrocarbon compounds, corresponds to a depth of more than 100 km. The results of the theoretical analysis shown in Fig. 2 clearly establish that the evolution of the molecular components of natural petroleum occur at depth *at least* as great as those of the mantle of the Earth, as shown graphically in Fig. 4 in which are represented the thermal and pressure lapse rates in the depths of the Earth.

As noted, the theoretical analyses reported in **section 5** describes the high-pressure evolution of hydrocarbons under the *most favorable* chemical conditions. The theoretical calculations for the evolution of hydrocarbons posited the presence of methane, the genesis of which must itself be demonstrated in the depths of the Earth consistent with the pressures required for the evolution of heavier hydrocarbons. Furthermore, the multicomponent system analyzed theoretically included no oxidizing reagents which would compete with hydrogen for both the carbon and any free hydrogen. The theoretical analysis assumed also the possibility of at least a metastable presence of hydrogen. Therefore, the theoretical results must be considered as the determination of *minimum* boundary conditions for the genesis of hydrocarbons. In short, the genesis of natural petroleum must occur at depths *not less than* approximately 100 km, - well into the mantle of the Earth. The experimental observations reported in **section 7** confirm theoretical predictions of **section 5**, and demonstrate how the iron compounds interact under high pressures to reduce water, of which the hydrogen combines with available carbon to produce heavy hydrocarbon compounds.

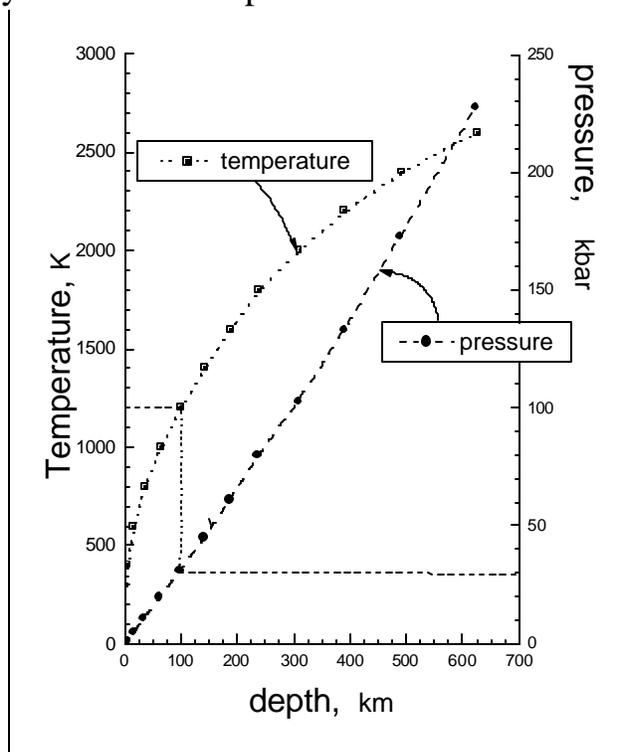

**Fig. 4 Pressure and temperature in the depths of the Earth.**

Notwithstanding the generality and first-principles rigor with which the present theoretical analysis has used, the results of the theoretical analyses here reported are robustly independent of the details of any reasonable mathematical model. The results of this theoretical analysis are strongly consistent with those developed previously by Chekaliuk, Stefanik, and Kenney, (27-30) using less accurate formal tools. The analysis of the H-C system at high pressures and temperatures has previously been impeded by the absence of reliable equations of state which could describe a chemically-reactive, multicomponent system at densities higher than such of its normal liquid state in ordinary laboratory conditions, and at high temperatures. The first analyses employed the (plainly inadequate) Tait equation(31); later was used the quantum mechanical Law of Corresponding States(32); more recently has been applied the single-fluid model of the SPHCT.(29, 30) Nonetheless, *all* analyses of the chemi-cal stability of the H-C system have shown results which are qualitatively identical and quantitatively very similar: all show that hydrocarbons heavier than methane cannot evolve spontaneously at pressures below 20-30 kbar.

**The H-C system does not spontaneously evolve heavy hydrocarbons at pressures less than approximately 30 kbar, even in the most favorable thermodynamic environment. The H-C system evolves hydrocarbons under pressures found in the mantle of the Earth, and at temperatures consistent with that environment.**

Significantly, these theoretical results are consistent with, and complement, the analysis of the genesis of the phenomenon of optical activity in abiotic fluids (including natural petroleum) previously reported.(14) In the past, observation of optical activity in natural petroleum had been spuriously claimed as evidence of a biological connection. Resolution of the problem of the intrinsic component of optical activity in natural petroleum has established that the imbalance of enantiomers which produces the effect is simply an inevitable manifestation of the complex behavior of multi-component systems composed of "closely-similar" molecules at high pressures.

In 1951, the Russian geologist Nikolai Kudryavtsev(33) enunciated what has become the modern Russian-Ukrainian theory of abyssal, abiotic petroleum origins, a fundamental tenet of which is that natural petroleum is a primordial, abiotic material, erupted from great depth. Kudryavtsev was soon joined by many prominent Russian geologists, geochemists, geophysicists, and petroleum engineers who together developed the extensive body of knowledge which now forms modern petroleum science.

Modern petroleum science has heretofore been a geologists' theory, supported by many observations, drawn into a comprehensive pattern, and argued by persuasion. By contrast, a physicist's theory uses only a minimum of data, applies fundamental physical laws, using the formalism of mathematics, and argues by compulsion. The theoretical results here reported, use only the fundamental laws of physics and thermodynamics, and establish the provenance of modern petroleum science in the rigorous mainstream of modern physics and chemistry. The experimental results here reported, confirm unequivocally those theoretical conclusions, which may now be taken as foundations of the modern theory of abyssal, abiotic petroleum origins.

**Dedication**:

In the first instance, this article is dedicated to the memory of Nikolai Alexandrovich Kudryavtsev, who first enunciated, in 1951(1), what has become the modern Russian-Ukrainian theory of abyssal, abiotic petroleum origins. After Kudryavtsev, all the rest followed.

This article is dedicated generally to the many geologists, geochemists, geophysicists, and petroleum engineers of the former U.S.S.R. who, during the past half century, developed modern petroleum science. By doing so, they raised their country from being, in 1946, a relatively petroleum-poor one, to the greatest petroleum producing and exporting nation in the world.

This article is dedicated specifically to the late Academician Emmanuil Bogdanovich Chekaliuk, the greatest statistical thermodynamicist ever to have turned his formidable intellect to the problem of petroleum genesis. In the Summer of 1976, during the depths of the cold war and at immeasurable hazard, Academician Chekaliuk chose to respond, across a gulf of political hostility, to an unsolicited letter from an unknown American chief executive officer of a petroleum company headquartered in Houston, Texas. Thenafter and for almost fifteen years, Academician Chekaliuk was my teacher, my collaborator, and my friend. [JFK]